\documentclass[journal=jprobs,manuscript=perspective,layout=twocolumn]{achemso}

\author{Bram Burger}
\email{bram.burger@uib.no}
\affiliation{Computational Biology Unit (CBU), Department of Informatics, University of Bergen, Norway}
\alsoaffiliation{Proteomics Unit (PROBE), Department of Biomedicine, University of Bergen, Norway}
\author{Marc Vaudel}
\email{marc.vaudel@uib.no}
\affiliation{Department of Clinical Sciences, University of Bergen, Norway}
\author{Harald Barsnes}
\affiliation{Proteomics Unit (PROBE), Department of Biomedicine, University of Bergen, Norway}
\alsoaffiliation{Computational Biology Unit (CBU), Department of Informatics, University of Bergen, Norway}

\title{On the importance of block randomisation when designing proteomics experiments}

\begin{document}

\begin{abstract}
  Randomisation is used in experimental design to reduce the prevalence of unanticipated confounders. Complete randomisation can however create unbalanced designs, for example, grouping all samples of the same condition in the same batch. Block randomisation is an approach that can prevent severe imbalances in sample allocation with respect to both known and unknown confounders. This feature provides the reader with an introduction to blocking and randomisation, insights into how to effectively organise samples during experimental design, with special considerations with respect to proteomics. 
\end{abstract}

\section{Introduction}

A vital part of experimental design consists of defining the order of sample processing and, if necessary, the creation of batches. The aim is then to avoid the introduction of confounders that would bias the interpretation of the data. One of the most famous examples of a confounded experiment is the observation of ``water memory'' by Davenas \latin{et al.} \cite{Davenas1988}, \latin{i.e.} the claimed ability of water to retain a memory of substances previously dissolved in it, which could not be replicated in a double-blinded experimental design \cite{Maddox1988}, suggesting that the initial data were the results of experimenter bias. Flaws in experimental design and unintended confounding can also be found in proteomics literature, as shown in \latin{e.g.} Sorace and Zhan \cite{Sorace2003}, Hu \latin{et al.} \cite{Hu2005}, Morris \latin{et al.} \cite{Morris2010}, and Mertens \cite{Mertens2016}. Knowing how to deal with the challenges of experimental design is therefore central to achieving reproducible experiments. For a general introduction to experimental design, see for example Box \latin{et al.} \cite{Box2005}, Ruxton and Colegrave \cite{Ruxton2006}, Lawson \cite{Lawson2015}, and for proteomics specifically, see Burzykowski \latin{et al.} \cite{Burzykowski2019} and Maes \latin{et al.} \cite{Maes2016}. 

While some common confounders, like sample annotation, date and order of processing, and their associated solutions are generic, others are more field-specific. In proteomics, the use of liquid-chromatography systems coupled to mass spectrometers (LC-MS) notably poses major challenges in terms of long-term performance and influence by outside factors \cite{Zhang2020}. Similarly, there can be differences introduced, notably during protein digestion \cite{Burkhart2012}, peptide fractionation and enrichment \cite{Solari2015}, and data interpretation \cite{Zhu2020}. A good experimental design must therefore take into account both generic and field-specific confounders, which can be extremely complex in larger experiments. This is especially challenging when experiments combine multiple analytical readouts, and when the allocation of patients to treatment arms and sample collection may be constrained. 

Given the constraints of an experiment, the goal of defining sample order and batches is to a) minimise the influence of anticipated confounders, b) mitigate the risk that unanticipated confounders bias the interpretation of the results, and c) ensure that the results remain interpretable if something does not go as planned and, for example, a batch is lost.

\section{Sample randomisation}

Deciphering the association between sample characteristics, \latin{e.g.} tumour types, and the proteome holds the key for improved diagnostic and treatment of diseases. In this setting, protein abundances are the responses, or outcome variables, of the experiment, and the other variables in the model are the explanatory variables \cite{Riter2005, Oberg2009, Maes2016}. Often, one is interested in studying the association of one or more explanatory variable(s) with the response(s), while having to control for other explanatory variables that are not of primary interest. The explanatory variables of interest are also referred to as treatment variables, \latin{e.g.} treatment, disease status, or tumour type. 

The variables included in the model because they are expected to potentially influence the outcome, although not being of primary interest, are referred to as control variables, and include properties such as enzyme batch, column, or day of acquisition. Note that some variables may belong in either category, depending on the goals of the experiment, \latin{e.g.} age, sex, and patient ancestry. If there is an association between treatment and control variables, this can impair the ability to estimate the effect of the treatment variable. In an extreme case, if \emph{Controls} are handled first, and \emph{Patients} last, to what extent are the observed differences between patients and controls genuine and not artefacts introduced during sample handling?

In addition to the monitored variables included in the model, other variables may also affect the results, such as machine drift or environmental changes during the analysis. Obviously one cannot control for all variables that may have an influence on the response variable. By their nature, unobserved variables cannot be included in the model as they are not observed, and including too many variables in the model reduces the power of the experiment. A high number of samples combined with randomisation provides a safeguard, in the long run, against undue influence of unobserved variables \cite{Senn1989, Senn2012, Box2005}. For more details on the importance of randomisation in proteomics, please see Morris \latin{et al.} \cite{Morris2010} and Mertens \cite{Mertens2016}. 

To illustrate the effects of randomised versus ordered allocation, let us assume that we have ten patients receiving a given \emph{Treatment} and ten a \emph{Placebo}, with the treatment resulting in a minor mean increase in the analytical readout --- \latin{e.g.} the  abundance of a given protein, see Figure 1. Figure 1A shows the experimental setting for the ordered allocation with \emph{Placebo} subjects processed first, while Figure 1E shows the same subjects in a complete randomised allocation. The `true' protein abundance for each patient is plotted in Figures 1B and F for the ordered and complete randomised allocations, respectively. Note that, apart from the order of the samples, these figures are exactly the same. 

Now let us introduce a machine drift (Figures 1C and G) that causes the mass spectrometer to detect slightly less of the protein over time. For the ordered allocation, the observed protein abundances show almost no difference between the two group means (Figure 1D). Conversely, the difference in group means for the randomised allocation is nearly the same as the `true' difference (Figure 1G), and only has added variance caused by the machine drift. Note that if the groups were reversed in the ordered sample allocation scheme, the group mean difference would have been exaggerated instead.

\begin{figure*}[h]
  \includegraphics[width=\textwidth]{Figure1rescale.pdf}
  \caption{Randomisation to account for machine drift. \textbf{A, E, I:} Sample allocation order for 10 subjects receiving placebo (black figures, subjects 1 -- 10) and treatment (red figures, subjects 11 -- 20); ordered allocation (A), complete randomisation (E), and block randomisation (I). \textbf{B, F, J:} True abundance for the subjects if there would be no machine drift; ordered allocation (B), complete randomisation (F), and block randomisation (J), (identical except for the sample order). \textbf{C, G, K:} Simulated machine drift, identical for all three settings. \textbf{D, H, L:} Observed abundances = true abundance + machine drift; ordered allocation (D), complete randomisation (H), and block randomisation (L). Dashed lines indicate the group means for each setting.}
\end{figure*}

\section{Block randomisation}

Complete randomisation can produce severely unbalanced sample allocations, \latin{e.g.} randomly assigning all subjects receiving treatment to one batch and all subjects receiving placebo to another batch. Then, batch and treatment are completely confounded, and it is impossible to perform an analysis with regards to the treatment. In such a situation it is not uncommon to resort to adjusting batches and sample ordering manually, or simply ``randomise until it looks good''. Both of these procedures are poorly reproducible, and potentially introduce unintended biases. A structured way to solve this problem is to rather rely on block randomisation \cite{Rosenberger2016}.

As a simple example, let us revisit the experimental setting from Figure 1. To ensure that both treatments are equally represented throughout the run, we make ten blocks of two subjects: one \emph{Treatment} and the other \emph{Placebo}. These are the smallest blocks we can make where each treatment is proportionally represented. The order of the treatments within the blocks (\emph{Treatment} first or \emph{Placebo} first) is chosen randomly for each block. Finally the subjects are randomly assigned to the blocks (see the right panels of Figure 1). Thus, we group small, representative subsets of the experiment together (the blocks), but within the blocks the order of the treatments is random, and within these constraints the assignment of subjects to blocks is also random. Consequently, while in complete randomisation all sample sequences are possible, block randomisation returns a sample sequence from a subset of all possible sequences where, by design, biases introduced by sequential processing are distributed as evenly as possible over the treatment groups.

In practice, it is not always possible or preferable to have the same number of subjects in all groups of the treatment variable(s) \cite{Senn1989, Senn2012}. The block randomisation procedure with groups of different sizes is only slightly different. As before, first, blocks of samples where each group is proportionally represented are created. For example, if one group is twice as large as the other group, each block would consist of three subjects, one of the smaller group and two of the larger group (Figure 2A). When the groups do not have a small common divisor, one can create blocks of different sizes. For example, in a nine vs. ten setting, one would make eight blocks consisting of one subject of each group, and one block with the remaining three subjects (Figure 2B). In an experiment with multiple treatment levels, \latin{e.g.} \emph{Placebo}, \emph{Treatment 1}, and \emph{Treatment 2}, the blocks would consist of subjects from all treatments. As previously, the blocks are put in random order, the order of the treatments within the blocks is chosen randomly for each block, and subjects are finally randomly allocated according to their characteristics.

\begin{figure}[h]
  \includegraphics[width=240pt]{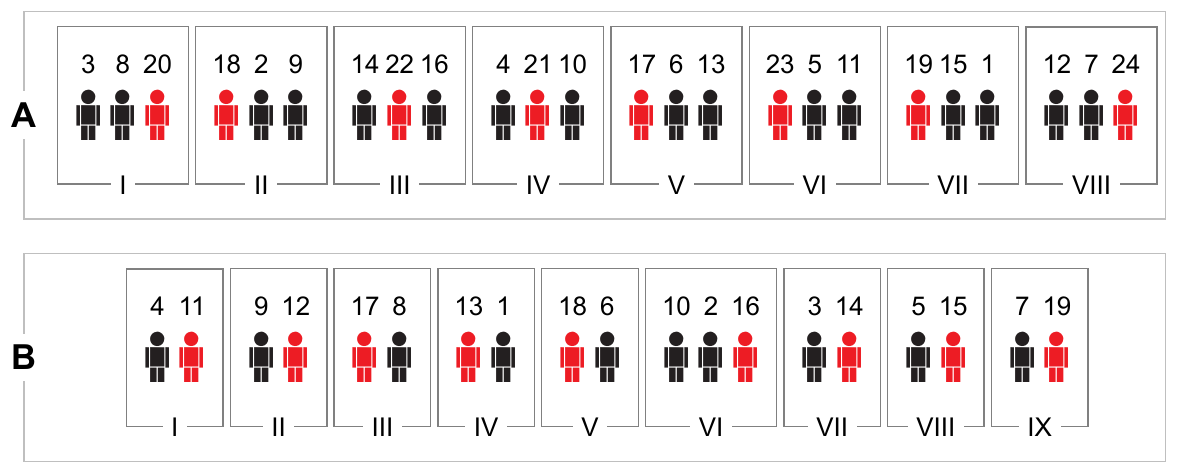}
  \caption{Examples of block randomisation. \textbf{A:} 16 subjects receiving placebo (black, subjects 1 -- 16) and eight treatment (red, subjects 17 -- 24), results in eight blocks of three subjects, each containing two \emph{Placebo} and one \emph{Treatment}. Subjects are randomly assigned to a block and the order within each block is randomised. \textbf{B:} 10 subjects receiving placebo (black, subjects 1 -- 10) and nine treatment (red, subjects 11 -- 19), results in one block of three subjects, containing two \emph{Placebo} and one \emph{Treatment}, and eight blocks of two subjects, containing one \emph{Placebo} and one \emph{Treatment}. The block containing three subjects is randomly placed among the other blocks. Subjects are randomly assigned to a block and the order of the subjects within each block is randomised.}
\end{figure}

\section{Accounting for control variables}

The previous examples included two groups of subjects where the treatment was assumed to be the only difference, and where all samples could be processed at the same time. Most experiments have to account for control variables when estimating the treatment effects. Some common control variables are technical in nature, such as protease batches, freezer locations, and biobanks, but sample characteristics such as sex, age, and patient ancestry also commonly fall into this category. As the complexity of the design increases, it is common that not all samples can be processed at the same time in the same way at the same location. The sets of samples created by this process are referred to as batches, and this becomes yet another control variable to account for.

It is important to distribute, as best one can, the different levels of the treatment variables equally over the different levels of the control variables. In case of substantial confounding, as when most of the subjects receiving \emph{Placebo} are \emph{Female} and most of those receiving the \emph{Treatment} are \emph{Male}, it can become impossible to estimate the treatment effect. Similarly, when all \emph{Treatment} subjects are in one batch and all \emph{Placebo} in the other, batch and treatment are confounded and without a batch-independent reference it is not possible to distinguish the treatment effect from a possible batch effect. Thus, it is important to account for control variables in the experimental design and in the sample organisation, and this can also be achieved using block randomisation.

As a simple example, let us reuse the experimental setting from Figure 1, but this time split the experiment into two batches, \latin{e.g.} due to different days of processing. Given that both treatment groups have ten subjects each, it is possible to divide both treatment groups into two subgroups of equal size. The first batch, \emph{Batch A}, consists of five randomly chosen subjects from the \emph{Treatment} group plus five randomly chosen subjects from the \emph{Placebo} group, while the second batch, \emph{Batch B}, consists of the remaining subjects (Figure 3A). The randomisation scheme is then exactly as before, except that half of the randomised blocks are now assigned to \emph{Batch A}, and the other half to \emph{Batch B}. 

Assume that for some unknown reason, something is different for the second batch, resulting in increased protein abundance measurements (Figure 3C). Each batch can now be seen as a separate experiment: the difference between treatment and placebo can be calculated in \emph{Batch A}, and similarly in \emph{Batch B} (Figure 3D). In both batches the treatment effect is close to the `true' difference (Figure 3B), even though the measured abundances are different.

\begin{figure}[h]
  \includegraphics[width=240pt]{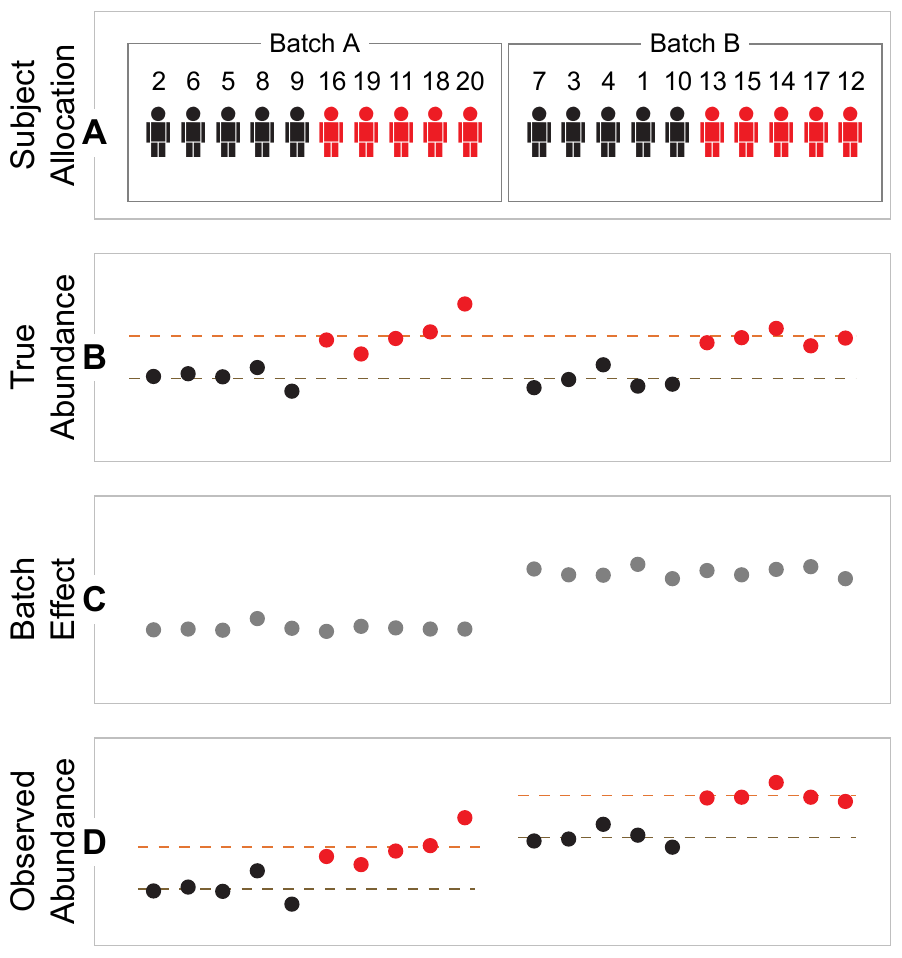}
  \caption{Randomisation to account for batches. \textbf{A:} Sample allocation order for subjects receiving placebo (black, subjects 1 -- 10) and treatment (red, subjects 11 -- 20). \textbf{B:} True abundance if no batch effect. Dashed lines indicate group means for all placebo subjects and all treatment subjects, considered to be the true group means. \textbf{C:} Simulated batch effect. \textbf{D:} Observed abundances = true abundance + batch effect. Dashed lines indicate group means within the batches.}
\end{figure}

When the experiment consists of multiple batches, they will undergo the same experimental protocol at different points in time and/or space. Every step of the protocol may then introduce variation that is specific for each batch. If samples are moved across batches between processing steps, each processing step has its own specific sample-to-batch allocation. Then, each processing step will have its own batch effect, and each of these will have to be estimated, unnecessarily increasing the complexity of the model. It is instead recommended to keep the same batches throughout the experiment, so that possible batch effects from different processing steps are combined into one overall batch effect. 

If different processing steps of the protocol have different size constraints, \latin{i.e.} one step requires more batches than another, being able to combine the smaller batches into larger batches without having to split the smaller batches is ideal. For example, when one experimental step can process 12 samples at once, while another step can process 24 samples at once, two batches from the first step can be combined for the second processing step. When this is not possible, it makes most sense to set up the batches according to the smallest constraints, and keep these batches throughout.

As a general advice, to reduce the complexity of experimental design, execution, and downstream analysis, it is recommended to keep the distribution of control variables as simple as possible. Note that this also applies to sample characteristics such as sex, age, and patient ancestry. For example, if one has to use two batches and is not interested in the association of the response with the sex of the patients, it is perfectly fine, and even recommended, to process \emph{Male} and \emph{Female} samples separately, hence confounding batch allocation with sex, thus reducing the complexity of the model.

\section{Multiple treatment variables}

When the effect of an explanatory variable, \latin{e.g.} sex, on the response is of interest, it must be considered as one of the treatment variables. Similar considerations then hold as for the control variables, \latin{i.e.} to be able to estimate the effect of the treatment variables, they should not be confounded with each other, nor with the control variables. 

In a setting with multiple treatment variables, one can also be interested in estimating interactions between the treatment variables. These interactions then become another treatment variable, and it is important to make sure that these are also not confounded with any of the other variables. Hence, with sex and treatment as the treatment variables, one has to make sure that the effects of both separately and their interaction can be estimated. For this, all combinations of sex and treatment should be present in the experiment. Additionally the comparison between these combinations should not be confounded by a control variable. Here again, the latter can be achieved using block randomisation.

For example, given 24 patients equally divided into two treatment levels, \emph{Placebo} and \emph{Treatment}, and two sexes, \emph{Female} and \emph{Male}, there are now four groups of six subjects representing each treatment-sex combination (1: \emph{Placebo Female}, 2: \emph{Placebo Male}, 3: \emph{Treatment Female}, 4: \emph{Treatment Male}). To make sure that the groups are balanced over the experimental design, we make six blocks of four subjects, one of each group, and within each block we randomly order the groups. Subsequently, we randomly allocate subjects to the blocks based on their group.

In case all samples cannot be processed together, one first creates the batches, and subsequently performs block randomisation within each batch. For example, say we have to process the above samples in two batches. Given that we are interested in all comparisons between the treatment-sex combinations, we make two batches of 12 samples, each batch containing three subjects of each group (Figure 4). This way, the batches are balanced in terms of the subject-characteristics that we use in the analytical model. 

\begin{figure}[h]
  \includegraphics[width=240pt]{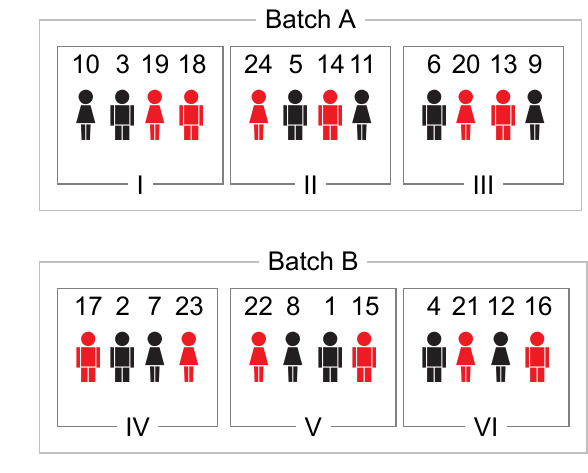}
  \caption{Example of block randomisation with two variables and two batches. Treatment variables are treatment and sex, with treatment having two levels: \emph{Placebo} (black) and \emph{Treatment} (red). All combinations of treatment variables are represented by one subject in each block. Within these constraints subjects are randomly assigned to a block and the order of the subjects in each block is randomised.}
\end{figure}

In contrast to the case where sex is a control variable, putting all \emph{Male}s in one batch and all \emph{Female}s in the other should now clearly be avoided: sex and batch would be  complete confounders, rendering impossible the comparison between \emph{Male}s and \emph{Female}s. Similarly, the interaction between sex and treatment would be lost. 

With an increasing number of variables, the model becomes increasingly constrained. Given that one always has only a limited number of samples, variables thus need to be prioritised. The general advice from Box \latin{et al.} \cite{Box2005} ``Block what you can and randomize what you can't'', implies that there is only so much one can control for. Especially in proteomics studies, with generally large heterogeneity between subjects, one has to be careful not to block so heavily, and to include so many variables, that some combinations of treatment and control variables occur in only a very limited subset of subjects.

\section{Further considerations}

The strategies outlined for the theoretical settings above can easily be extended to more elaborate situations. However, as for all methods, implementing block randomisation can quickly become challenging in real-world situations. In this section, special considerations are introduced for situations where the reality of the experiment poses challenges in experimental design.

\subsection{Continuous variables and fuzzy categories}

Most of the variables inspected so far have divided the samples into non-overlapping categories, this will however not always be the case. Often, variables are continuous or categories are overlapping, such as age or disease state, respectively. In both cases categorisation is commonplace, but can be problematic. Categories that span a large number of values can lead to relatively large differences between subjects within a category, while the differences between subjects at the edges of neighbouring categories will be small. On the other hand, in a large enough study the randomisation should mitigate this problem. Additionally, a substantial number of subjects per category is a requirement to still be able to randomise subjects. If each subject ends up being its own category, randomisation will no longer be possible. 

In cases where categorisation is problematic, one can perform matching for the purposes of blocking, \latin{i.e.} subjects that are similar according to a given variable \latin{e.g.} age or disease state, but belong to different groups with respect to all other variables, are treated as a single group for the purposes of block randomisation (see for example Sekhon \cite{Sekhon2011}). Note that the categorised version is solely created for block allocation, and the final analysis still uses the original variable.

\subsection{When batch-size is smaller than the number of groups}

When the number of variables increases, the batch-size may become smaller than the number of groups. It then becomes important to distribute samples across batches in a way that makes it possible to calculate the differences of interest. This generally means that one has to prioritise some group-comparisons over others. The more often a pair of groups occurs together in a batch, the better one can estimate the given difference. On the other hand, when a pair of groups never occurs together in a batch, but share a batch with a common other group, the difference can be calculated through this common other group. For general strategies on how to construct batches in a way that the differences of interest are estimable, see \latin{e.g.} Lawson \cite{Lawson2015} or Box \latin{et al.} \cite{Box2005}.

\subsection{Mitigation of batch effects using reference samples}

An approach that maintains comparability between samples is to introduce a common reference. For example, in targeted proteomic analyses, one spikes a known amount of the heavy version of a peptide/protein, for use as a reference to infer the concentration of the peptide/protein present in each sample, and hence compare the abundance of the specific peptide/protein across the samples \cite{Maes2016,Kuzyk2009}.

A similar approach can also be used in untargeted settings, both with and without labels \cite{Geiger2010, Zhang2020, Brenes2019}, where one sample is used as a standard throughout the experiment. Having a common reference makes samples more easily comparable across the different settings (\latin{e.g.} batches, days of analysis, and instruments) by providing a common baseline. However, this only works for processing steps that the reference sample shares with the other samples in the relevant batch, and poses challenges in terms of missing value and dynamic range that are beyond the scope of this article. 

The use of common reference samples can alleviate many challenges concerning batch effects. However, it is not always possible or desirable to include a common reference. For example, there may be constraints with regards to the available resources, \latin{e.g.} the acquisition of heavy peptides for absolute quantification in targeted studies can be very expensive, and similarly, for large studies that run for an extended period of time, it is often not feasible to create a reference sample with a comparable composition relative to the experimental samples and that is large enough to last through the entire study.

\section{Closing remarks}

Proteomics has many aspects that ought to be taken into account when designing and planning experiments. The complexity of the samples, the proteome, and the analytical techniques employed make proteomics experiments particularly challenging. Especially in larger studies, the labour intensive sample preparation often means that the experiment has to be split into multiple batches. It is therefore important to design the experiment in such a way that variables and batches are not confounding. Each batch should as much as possible be its own small experiment. When the size of the batches is insufficient to achieve this, it is essential to make sure that the utilised experimental design can give answers to the scientific questions asked. To ensure this, and minimise the bias due to unobserved variables as much as possible, one should therefore block variables and randomise subjects, in the different batches, and use this restricted random allocation to process samples in the lab and on the mass spectrometer.

It is important to underline that the challenges posed by the handling of multiple variables can only be answered if the scientific project is rigorously defined, with response, treatment, and control variables clearly identified before the samples are collected. Given the multidimensional and multidisciplinary nature of modern omics projects, it is essential that experts with the necessary expertise are involved early in the experimental design, to prevent confounding effects. Finally, while considerations of power are beyond the scope of this article, it cannot be stressed enough that an adequate number of samples is paramount, both for correct experimental design and to ensure that the research questions can be answered.

\bibliography{references}
\end{document}